\documentclass[twocolumn,showpacs,preprintnumbers,amsmath,amssymb]{revtex4}

\usepackage{graphicx}% Include figure files
\usepackage{dcolumn}% Align table columns on decimal point
\usepackage{bm}% bold math

\begin{document}
% \draft command makes pacs numbers print

%\preprint{03.75.-b, 39.20+q, 42.50.Vk}

\title{Interferometer-Type Structures for Guided Atoms}
%\\
%based on Microfabricated Optical Elements}

\author{R. Dumke, T. M\"uther, M. Volk, W. Ertmer,
and G. Birkl\footnote{Author to whom correspondence should be addressed.}}

\affiliation{Institut f\"ur Quantenoptik, Universit\"at Hannover,
Welfengarten 1, D-30167 Hannover, Germany}

\date{\today}

\begin{abstract}
We experimentally demonstrate interferometer-type guiding
structures for neutral atoms based on dipole potentials created by
micro-fabricated optical systems. As a central element we use
an array of atom waveguides being formed by focusing a red-detuned
laser beam with an array of cylindrical microlenses. Combining two
of these arrays, we realize X-shaped beam splitters and more
complex systems like the geometries for Mach-Zehnder and
Michelson-type interferometers for atoms.
\end{abstract}

% insert suggested PACS numbers in braces on next line

\maketitle

%\narrowtext

%%%%%%%%%%%%%%%%%%%%%%%%%%%%%%%%%%%%%%%%%%%%%%%%%%%%%%%%%%%%%%%%%%%%%%%%%%%%%%

The investigation and exploitation of the wave properties of
atomic matter is of great interest for fundamental as well as
applied research and therefore constitutes one of the most active
areas of research 
in atomic physics and quantum optics. Of special interest is
the field of atom interferometry \cite{Atominterferometry}. In
comparison to optical interferometers, atom interferometers have
the potential of being several orders of magnitude more sensitive
for some applications or giving access
to classes of interferometric measurements not being possible
with optical interferometry in principle. 
%(such as the measurement of gravity).
In the last decade, an impressive list of high-precision atom
interferometrical measurements of e.g. fundamental constants, 
atomic properties, inertial forces,
and rotations  have been performed
\cite{Sterr,Young,Ekstrom,Gibble,Clairon,Peters,
Riehle,Lenef,Gustavson}.

Because of the high intrinsic sensitivity, these interferometers
have to be built in a robust way to be applicable under a wide
range of environmental conditions.
%This can be achieved by guiding the atoms along a potential, and
%integrate this setup in a miniaturized device.
A new approach to meet
this challenge lies in the development of miniaturized and
integrated atom optical setups based on micro-fabricated
guiding structures.
%\cite{Weinstein,Hindsreview,Schmiedmayer,Haensch,Cornell,Prentiss,Engels,Hinds}.
%The field of atom
%optics has been established and research in this area has already
%led to many exciting results. Many of the future applications of
%cold atomic ensembles rely on the development of efficient atom
%optical structures for atomic matter waves.
Using micro-fabricated current carrying wires, 
several configurations for atom guides
\cite{Dekker,Müller99,Key00,Folman00,Engels,Sidorov} and beam splitters
\cite{Müller01,Cassettari00,Hänsel01} 
also using Bose-Einstein condensates \cite{MicroBEC}
have been realized.
As an important goal remains the demonstration of a
setup suitable as a guided-atom interferometer.
%A further
%investigation is the development of compact interferometrical
%devices. Such a device based on atomic matter waves is more
%precise than the light interferometers in several areas
%\cite{Atominterferometry}.

\begin{figure}

  \includegraphics{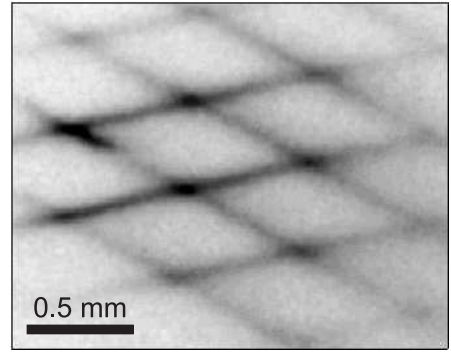}

\caption{\label{machc} Fluorescence image of rubidium atoms
loaded from a MOT into a multiple-path Mach-Zehnder-type guiding structure
created by the combined
dipole potentials of two sets of micro-fabricated arrays of
cylindrical lenses.}
\end{figure}

In this letter we present the experimental implementation of
atom guides, beam splitters, and structures for atom
interferometers based on
micro-fabricated optical elements 
as proposed in \cite{unseroptcomm}.
We demonstrate the guiding of neutral atoms along
focal lines of arrays of micro-fabricated cylindrical lenses
making use of optical dipole potentials. By
superimposing two of these arrays under a variable relative angle,
we realize X-shape beam splitters as well as interferometer-type
configurations like Mach-Zehnder (Fig. \ref{machc}) or
Michelson-type structures. 
Due to the state-selectivity of optical
potentials, a state-selective guided-atom beam splitter could be
demonstrated as well. Theoretical simulations of the wave dynamics
predict the required coherence and
interferometrical properties of these configurations
\cite{kreutzmann}(see also \cite{Andersson02}).
%Together with systems based on miniaturized and micro fabricated
%electrostatic and magnetic devices, the application of
%microoptical systems will launch a new field in atom optics which
%we call ATOMICS for Atom Optics with MICro Structures.
%

For the experiments presented here, we employ one-dimensional
arrays of cylindrical microlenses. 
Laser light sent through such a system forms a series of
parallel focal lines above the lens array. 
Thus, for laser light detuned below an atomic
resonance ('red-detuning') a one-dimensional array of atomic
waveguides is formed (Fig. \ref{velocity} (a)) \cite{uniformity}.
Atoms are confined in the two dimensions
perpendicular to the lens axis but are free to propagate along the
longitudinal axis for a homogeneous intensity distribution along
the lens axis. 
%On the other hand, 
%the motion of the atoms can be controlled in a
%simple fashion by altering the intensity distribution and
%thus the shape of the potential along the
%longitudinal direction.
Besides creating longitudinally flat potentials we can apply more
complex intensity distributions, such as an intensity gradient for
accelerating the atoms.
%how the atoms were loaded at
%the slope of flat potential and a curved potential where they were
%accelerate along the guide due to the gradient of the potential.
%The mean velocity is increasing with time.
A light field with a gaussian intensity profile along the axis of
the guide allows us to reverse the atom motion in the guide.
This design flexibility gives the possibility to integrate atom
mirrors with guiding structures in a direct fashion.

In our setup, the lens array consists of 12 lenses with a
length of 5 mm, manufactured in a fused sillica substrate.
The center to center separation as well as the
diameter of each lens are 0.4 mm.
% witch is also the diameter of each lens.
The focal length is 2.21 mm giving a numerical aperture 
NA = 0.09. 
We image the focal plane of the microlens
array onto the atoms to be guided with the help of two achromats
(NA = 0.08, magnification = 1). The
optical transfer of the trapping light has the
advantage that we can place the microoptical systems
outside the vacuum chamber and thus can switch between and
superimpose several microoptical elements easily ({Fig.
\ref{velocity} (b)). The guiding of atoms close to the surface
of these and of more complex integrated 
structures \cite{unseroptcomm,Burke02}
can be achieved in a straightforward fashion
by putting the elements directly inside the
vacuum chamber.

A typical experimental sequence starts by loading a single dipole
trap with about $10^{4}$ $^{85}Rb$ atoms 
at a measured temperature of 20 $\mu$K from a MOT
(see
\cite{Dumk01} for details).
% therefore only the first 50 vibrational
%levels should be occupied on average.
After loading, the atoms are held in the single dipole trap for 35
ms, for untrapped atoms being able to leave the detection region.
Then we transfer the atoms into the guiding
structures by turning on the guiding and turning off the
single-dipole trapping light.
The loading efficiency is close to unity
because the single dipole trap is formed by illuminating a small
part of the same cylindrical microlens, thus the reloading into the
guide is achieved without significant loss in atom number or
increase in temperature. Taking advantage of this, it should be
possible to prepare the atoms by e.g. Raman sideband cooling
\cite{Vuletic98} in the ground state of the single dipole trap and
reload them into the guiding structure adiabatically without
changing the transversal vibrational state.
The temporal evolution of the atom
distribution in the guiding structures is observed via
fluorescence imaging by a CCD camera with a
spatial resolution of 14 $\mu$m (rms-spread). For detection, the
guides are switched off and the atoms are illuminated by the MOT
light for a period of 0.8 ms.

\begin{figure}

  \includegraphics{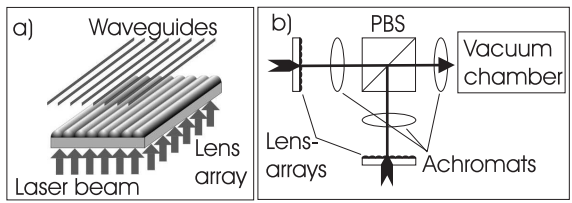}

\caption{\label{velocity}a) Array of atom waveguides formed by
focusing a red-detuned laser beam with an array of cylindrical
microlenses. b) Schematic of the optical setup used for combining
and transfering light fields: The focal planes
of two lens arrays are reimaged by achromats and combined by a
polarizing beam splitter (PBS).}
\end{figure}

The light used for atom guiding is derived from a single-frequency
Ti:Sapphire laser and has a typical power of 360 mW and a typical 
detuning $\Delta\lambda$ = 1 nm 
($\delta\nu$ = -500
GHz $\approx -5 \times 10^{5}$ linewidths) 'red' of the $5S_{1/2}
\rightarrow 5P_{3/2}$ transition at 780 nm. A typical
potential depth for the waveguides is
450 $\mu$K. The radial (i.e. perpendicular to the laser beam 
direction) waist is 7 $\mu$m ($1/e^2$
radius of the intensity distribution) and the corresponding
Rayleigh range is 200 $\mu$m. 
The calculated radial oscillation
frequency is 9.6 kHz. 
From this and the measured temperature of 20 $\mu$K
we infer a mean occupation number
of $<n>\approx 40$ (radial direction) and a rms position spread of 
$\approx$ 1 $\mu$m in radial direction and of $\approx$ 5 $\mu$m
along the laser beam direction.
The rate for spontaneously scattered
photons is 740 $s^{-1}$ \cite{besser}.
%Mention that the scattering time is not equal to the coherence
%time for the external degrees of freedom \cite{Buchkremer00}.
With this configuration we could observe guiding of atoms over a typical
distance of 2.5 mm, limited by the accessible length of the
lens array. 

\begin{figure}

  \includegraphics{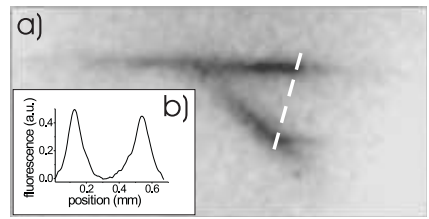}

\caption{\label{Beamsplitter} Beam splitter for guided atoms:
a) Fluorescence image of atoms guided through a beam splitter based on two
crossed cylindrical microlenses. b) Line profile of the atom distribution 
along the dotted line of (a).}
\end{figure}

An important element for guided-atom interferometry is a beam
splitter for atoms. Such a structure can be realized by combining
the light fields of two microlens waveguides having one
common focal plane but being oriented non-parallel to each other
(see also \cite{Houde00}).
The combination of the two waveguides results in
an X-shaped beam splitter as shown in Fig. \ref{Beamsplitter}. 
Atoms are loaded from a single dipole trap into the input port
at left and are accelerated towards 
the intersection region by a gradient in the guiding potential.
Here, the paths split and the atoms are guided along both output ports.
By rotating
one lens array with respect to the other, the splitting angle can
be chosen arbitrarily. For our experiments, it was set to
$42^{\circ}$ degrees.

Interference effects in the combined light field at the intersections where 
the individual potentials add, are avoided by polarizing the
two light fields orthogonally (see {Fig. \ref{velocity} (b)).
The
transverse mode structure of the input port is identical to the
one of the output ports with the potential being twice as deep at
the intersection. 
Detailed calculations based on the parameters of
our configuration show that the splitting process indeed is
coherent for the guiding potential evolving adiabatically
during the splitting process \cite{kreutzmann}.
The highest possible degree of coherence can be achieved by
compensating the doubling of the potential depth due to the
overlap of the trapping potentials at the intersection by an
optimized lens design \cite{unseroptcomm} or by overlapping the 
intersection region with the repulsive potential of 
a blue-detuned laser beam \cite{calculation}.

We could demonstrate a variation in the splitting ratio over a
wide range by changing the power ratio between the two guides
forming the beam splitter
(Fig. \ref{Powerratio}).
The atom number in the two output ports is determined
by taking line profiles similar to 
Fig. \ref{Beamsplitter} (b) and integrating the density distribution for
each port. By choosing an appropriate power ratio, which depends
on the velocity of the atoms and the splitting angle, a splitting
ratio of 50/50 can be achieved.

\begin{figure}

  \includegraphics{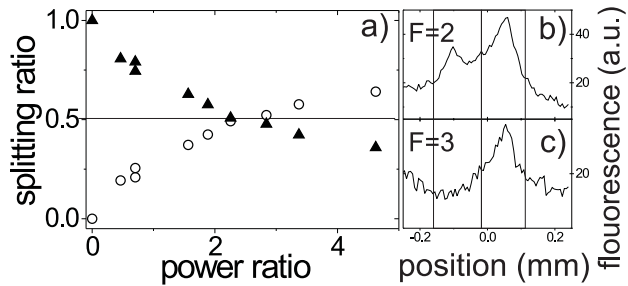}

\caption{\label{Powerratio} Properties of different types of
beam splitters: a) The splitting ratio for atoms in a
state-insensitive beam splitter is altered by varying the
intensity ratio between the waveguides creating the beam splitter. 
b) and c) Line profiles
for atoms after moving through a state-selective beam splitter:
b) Atoms in the state
$5S_{1/2} (F=2)$ evolve through the beam splitter unperturbed. The
line profile shows atoms in both outputs. c)
Atoms in the state $5S_{1/2} (F=3)$ can only propagate
along one output.}
\end{figure}

So far, the splitting process was fully determined by the external
degrees of freedom. 
Specific to optical guiding structures is the possibility
to use the
internal atomic structure for the splitting process, similar to
Ramsey-Bord\'{e} \cite{Riehle,Sterr} and Raman interferometers
\cite{Young} for free atoms. 
For guided atoms,
state-selective splitting can be achieved for example by applying an additional
state-selective optical potential
in a small
section of one output shortly behind the
beam splitter of Fig. \ref{Beamsplitter}.
We implemented this
 %a state-selective guided-atom
%beam splitter 
for $^{85}Rb$ by employing an additional laser field with a
red-detuning of $\delta \nu_{2} = -1020$ MHz (-170 linewidths) for
atoms in the $5S_{1/2} (F=2)$ hyperfine ground state.
%\rightarrow 5P_{3/2} (F'=3)$ transition. 
The
same laser field is blue-detuned for atoms in the 
$5S_{1/2} (F=3)$ hyperfine ground state
with a detuning of 
$\delta \nu_{3}=2020$ MHz (340 linewidths)
%
%$\Delta \nu_{3}=\Delta_{HFS} - \Delta \nu_{2}$
%with respect to the $5S_{1/2} (F=2) \rightarrow 5P_{3/2} (F'=3)$
%transition (with $\Delta_{HFS}$ being the hyperfine splitting of
%the ground state) 
%
\cite{Stateselective}. 
State-selective splitting
for a guided-atom beam splitter is shown in Fig. \ref{Powerratio}
(b) and (c). 
For atoms in the $5S_{1/2} (F=2)$ state, 
%propagating through a beam splitter similar
%to the one of (Fig. \ref{Beamsplitter}),
after the intersection there is one
unperturbed output and one with an additional potential well (simply acting as a
phase shifter). Therefore the atoms propagate along both outputs
of the splitter (Fig. \ref{Powerratio} (b)). Atoms initially prepared
in the $5S_{1/2} (F=3)$ state cannot propagate
along the output with the additional potential barrier caused by
the blue-detuned light field. (Fig. \ref{Powerratio} (c)). A
foreseeable application of this technique will lead to the
preparation of atoms in a superposition of the two hyperfine
ground states prior to or right at the beam splitter, so that the
splitting process will be determined by the internal superposition
state. With such a system it should be possible to create robust coherent
beam splitters for guided atoms based on their internal states.

As being the central goal of this work, 
we designed and experimentally demonstrated the structures for two
integrated interferometer-type configurations for guided atoms:
(i) Mach-Zender-type (Figs. \ref{machc} and \ref{mach}) 
and (ii) Michelson-type (Fig. \ref{Michel})
structures.

\begin{figure}

  \includegraphics{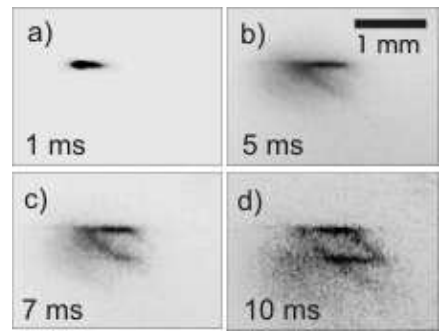}

\caption{\label{mach} Temporal evolution of atoms propagating through a
Mach-Zehnder-type interferometer structure: Loaded from a single
dipole trap (a) the atoms are split by the input beam splitter (b) and
propagate along two different paths to the
output beam splitter (c, d).}
\end{figure}

By combining two arrays of waveguides we create
multiple X-shaped beam splitters. A set of four of these beam
splitters represents the basic Mach-Zehnder-type structure (Fig. \ref{mach})
with the
option of using the additional input and output ports for creating more
complex multi-path guided-atom interferometers (Fig. \ref{machc})
e.g. for increased sensitivity \cite{multi}.
The basic configuration acts as a closed loop structure with a
finite enclosed area as required for a Sagnac-type
interferometer.
Fig. \ref{mach} shows the propagation of atoms through a
Mach-Zehnder-type structure for guided atoms. We load one of the
input ports with atoms from the single dipole trap (a). Atoms
propagate to the first beam splitter and split into two paths
(b). At the next intersections these split
into a total of four paths (c). Two of the paths recombine at the the
fourth intersection (d).
This Mach-Zehnder-type structure has an enclosed area of 0.3
$mm^{2}$ with a total required array
below 1 $mm^{2}$ including the loading and 
detection regions.
It presents the first experimental demonstration of a
closed structure suitable for atom interferometry based on atom
guides. In numerical simulations, it could be demonstrated that
for typical experimental conditions coherent splitting of
atom-waves and matter-wave interference at the outputs can be achieved
\cite{kreutzmann}. As an important result, a variation in the
relative phase between the two paths of a Mach-Zehnder-type structure
resulted in a complementary
periodic variation of atom number in the two final output ports,
thus clearly predicting the existence of interference fringes.

We also demonstrated a Michelson-type interferometer
structure (Fig. \ref{Michel}) 
by crossing two microoptical guides in our 'standard'
beam splitter configuration, each guide now having a gaussian
intensity profile along the longitudinal direction 
centered at the intersection (Fig. \ref{Michel} (a)).
%Atoms loaded on one side of the intersection
%cross the beam splitter and are reflected by the increasing
%potential on the other side.
%We experimentally demonstrated this Michelson-type structure
%as well.
We load atoms in one input close to the edge of the potential
well. The atoms are accelerated towards the beam splitter (Fig.
\ref{Michel} (b)). They split into two paths, and are slowed
down by the positive gradient of the potential (Fig. \ref{Michel}
(c)). They start moving backwards towards the intersection. At the
intersection they split again with part of the atoms now
also moving along the output port of the Michelson-type structure (Fig.
\ref{Michel} (d)). The oscillation along the longitudinal
direction is significantly slower than in the transverse directions. 
Thus, an
adiabatic propagation through the structure is easily achievable.

\begin{figure}

  \includegraphics{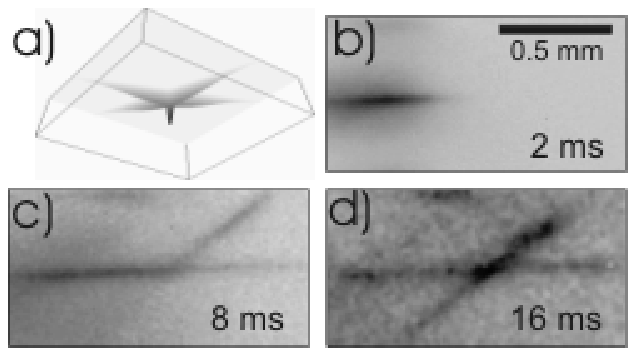}

\caption{\label{Michel} Configuration of a Michelson-type
interferometer: a) Structure based on two crossed waveguides with a
gaussian intensity profile along the longitudinal direction. b) - d)
Propagaton of atoms in the Michelson-type structure loaded
from a single dipole trap.}
\end{figure}

%
%By
%realizing such a type of atomic interferometer it should be
%possible to measure accelerations, or the gravity in a high
%precision.
%
In this letter we have presented the first experimental
demonstration of structures for guided-atom
interferometers based on micro-fabricated elements. For this
purpose, guides and beam splitters based on microoptical elements
have been employed. Theoretical investigations predict the
feasability of coherent beam splitting and the possibility of
achieving matter wave interference for guided atoms. 
However, several modifications to our apparatus, such
as improved atomic state preparation, detection resolution, and detection
efficiency need to be implemented prior to a demonstration of coherence. 
Interferometer structures 
optimized for coherence and fringe contrast are currently
being developed and the demonstration of an interference 
experiment will be attempted under optimized experimental
conditions.
The
microoptical systems investigated here are first realizations of a
broad class of configuration being accessible due to the high
flexibility in the manufacturing process of these elements
\cite{unseroptcomm}. Applications hugely benefit from the many
inherent advantages of integrated systems, such as stability and
scalability. The configurations described here present a major step
towards miniaturization of atom-interferometrical devices and
promote the wide practical applicability of various classes of sensors based on atom
interferometry.

We thank F. Buchkremer, H. Kreutzmann, M. Lewenstein, and A. Sanpera 
for many stimulating discussions. This work is supported by the
SFB 407 
and the SPP {\it Quanteninformationsverarbeitung}
of the {\it Deutsche Forschungsgemeinschaft} and by the
project ACQUIRE (IST-1999-11055) of the {\it European Commission}.

%%%%%%%%%%%%%%%%%%%%%%%%%%%%%%%%%%%%%%%
%\begin{thebibliography}

%\end{thebibliography}


\begin{references}
\bibitem{besser}
Lens arrays with optimized numerical
aperture will give further improved waveguide parameters 
(see \cite{unseroptcomm})
\bibitem{calculation} The calculations 
of \cite{kreutzmann} show that
significant coherence is still maintained during the splitting process
with a mismatch in the potential compensation of up to 20$\%$.
\bibitem{Stateselective} Other implementations of a state-selective
 beam splitter can be based on the polarization dependence of the optical
 potentials for different internal atomic states.
\end{references}
\end{document}